\documentclass[twocolumn,prb,showpacs,superscriptaddress,amsmath,amssymb]{revtex4}
\usepackage{graphicx}
\usepackage{dcolumn}
\usepackage{bm}
\usepackage{color}
\begin{document}

\title{Suppression of asymmetric differential resistance in non-Fermi liquid system YbCu$_{5-x}$Al$_{x}$ (x = 1.3 - 1.75) in high magnetic fields}

\author{G. Prist\'a\v s}
 \email{gabriel.pristas@saske.sk}
\author{M. Reiffers}
\affiliation{Institute of Experimental Physics, Slovak Academy of Sciences, Watsonova 47, 040 01 Ko\v sice, Slovakia}

\author{E. Bauer}
 \affiliation{Institut f\H ur Festk\H operphysik, Technische Universit\"at Wien, 1040 Wien, Austria}

\author{A. G. M. Jansen}
 \affiliation{Service de Physique Statistique, Magn\'etisme, et Supraconductivit\'e, Institut Nanosciences et Cryog\'{e}nie
 CEA-Grenoble, 38054 Grenoble Cedex, France }

 \author{D. K. Maude}
 \affiliation{Grenoble High Magnetic Field Laboratory (CNRS), BP 166, 38042 Grenoble Cedex 9, France}%

\date{\today}

\begin{abstract}
The non-Fermi liquid system (NFL) YbCu$_{5-x}$Al$_{x}$ ($x$ = 1.3 -
1.75) has been investigated in hetero- as well as homo-contact arrangement
in magnetic fields up to 22.5~T. The observed
d$V$/d$I$(V) characteristics reveal asymmetry in hetero-contact arrangement and do not agree with the model of thermal
contact heating, at least close to zero-bias voltage. In the case of
a hetero-contact arrangement we have observed a maximum at only one
voltage polarity at about 1.3 mV (for $x$ = 1.5), which asymmetry is
suppressed in an applied magnetic field. This behavior is connected
with NFL properties of studied compounds.
\end{abstract}

\maketitle

\section{INTRODUCTION}

Point-contact spectroscopy (PCS) is a very efficient tool for the
study of electronic scattering processes in metallic conductors.
\cite{Naidyuk} In the case of ballistic transport of the charge
carriers across the contact, the applied contact voltage defines
directly the energy scale of the scattering processes investigated,
e.g. the electron-phonon interaction. This is valid for simple metals and compounds, but there are some polemics about applicability of PCS in more complicated systems, like heavy-femion systems a systems with strongly correlated electrons. The most of papers about PCS and tunneling spectroscopy are connected with superconducting systems, where the critical points are covered by superconductivity.
Here, we present the results of the first application of PCS to the non-superconducting system
YbCu$_{5-x}$Al$_{x}$ ($x = 1.3 - 1.75$) showing non-Fermi liquid
(NFL) behavior in the vicinity of quantum critical point (QCP).

YbCu$_{5-x}$Al$_x$ is a very interesting compound with a valency
change from $\nu \approx$ 2.2 ($x = 0$) to $\nu \approx$ 3 ($x =
2$), which causes a magnetic instability near a critical
concentration $x_{cr} = 1.5$ in the cross-over from the almost
nonmagnetic $4f^{14}$ state in YbCu$_{5}$ to the magnetic $4f^{13}$
state in YbCu$_{3}$Al$_{2}$.\cite{Bauer} The intermetallic system
YbCu$_{5-x}$Al$_{x}$ with concentration in the vicinity of $x_{cr}$
exhibits a typical NFL behavior, like a negative logarithmic term in
the temperature-dependent specific heat or deviations from the
quadratic temperature dependence of the electrical
resistivity.\cite{Bauer} Such a NFL behavior is frequently found in
strongly correlated electron systems where the
magnetic ordering temperature tends towards zero leading to a QCP.
In our system the QCP is near $x_{cr}$.\cite{Bauer} There exist
scenarios for the occurrence of NFL behavior,\cite{Coleman, Stewart}
but the microscopic basis of the NFL ground state is not yet
completely understood. Theoretical work by Shaginyan and Popov\cite{Shaginyan2007} predict the asymmetric shape of dynamic conductance for strongly correlated electron systems in the vicinity of QCP in normal and/or superconducting state. This asymmetric part is expected to have
the linear contribution (part). Moreover, the absence of the classical
quasiparticles in spectra is characteristic for NFL systems.\cite{Shaginyan2007}

Our previous PC experiments on YbCu$_{5-x}$Al$_{x}$
 ($x = 1.3 - 1.6$) have been preformed at temperatures down
to 1.5 K and in magnetic fields up to 6 T in the hetero-contact
configuration using a Cu or Pt
counter-electrode.\cite{Pristas, Reiffers} The differential
resistance d$V$/d$I$(V) as a function of the applied voltage V
revealed an asymmetric behavior. In the case of $x_{cr} = 1.5$ we
have observed a maximum at 1.3 mV in only one voltage polarity. This
new type of asymmetry is connected with the NFL
behavior at the QCP.\cite{Pristas, Reiffers} The application of a
magnetic field strongly changed the shape of the differential
resistance, showing a recovering of the Fermi-liquid (FL) behavior
characteristic for Kondo compounds.\cite{Reiffers1992, Duif1989} The
behavior of PC dependencies of other concentrations differ from the
one at the critical concentration.

Since previous measurements were done in the magnetic fields up to 6
T only, which is not enough to fully suppress the NFL behavior and
restore the FL behavior, we applied high magnetic fields up to 22 T.
In this overview we present systematic point-contact measurements in
the hetero- and homo-contact arrangement at low temperatures in
magnetic fields up to 22 T, studying several YbCu$_{5-x}$Al${_x}$
compounds in the vicinity of QCP ($x_{cr}$ = 1.5). Measurements of
hetero-contacts were done down to 100 mK. We used the same samples
as characterized and studied in previous work of E.
Bauer \textit{et al}.\cite{Bauer, Bauer2}

\section{EXPERIMENTAL DETAILS}
Samples have been prepared from stoichiometric amounts of elements using high frequency melting.
The ingots were remelted several times and subsequently annealed
for 10 days at 750 $^{\circ}$C. The detailed fabrication of samples is described in
Bauer \textit{et al}.\cite{Bauer2}

Measurements were carried out by the 'needle-anvil' technique
\cite{Jansen, Yanson} using a differential screw mechanism for the
tip-sample adjustment. Before each series of measurements the
polycrystalline sample was mechanically polished or freshly broken
in order to obtain a clean surface. In the case of hetero-contacts,
the counter-electrode was made from a copper or a platinum wire of
100 $\mu$m diameter with a sharpened tip. In the homo-contact
arrangement, both parts of the contact were made from broken pieces
of the polycrystalline sample under study. We recorded the
differential resistance d$V$/d$I$(V) of the current-voltage
characteristic using standard phase-sensitive
detection.\cite{Jansen, Yanson}

\begin{figure}
\includegraphics [angle=270, width=80mm]{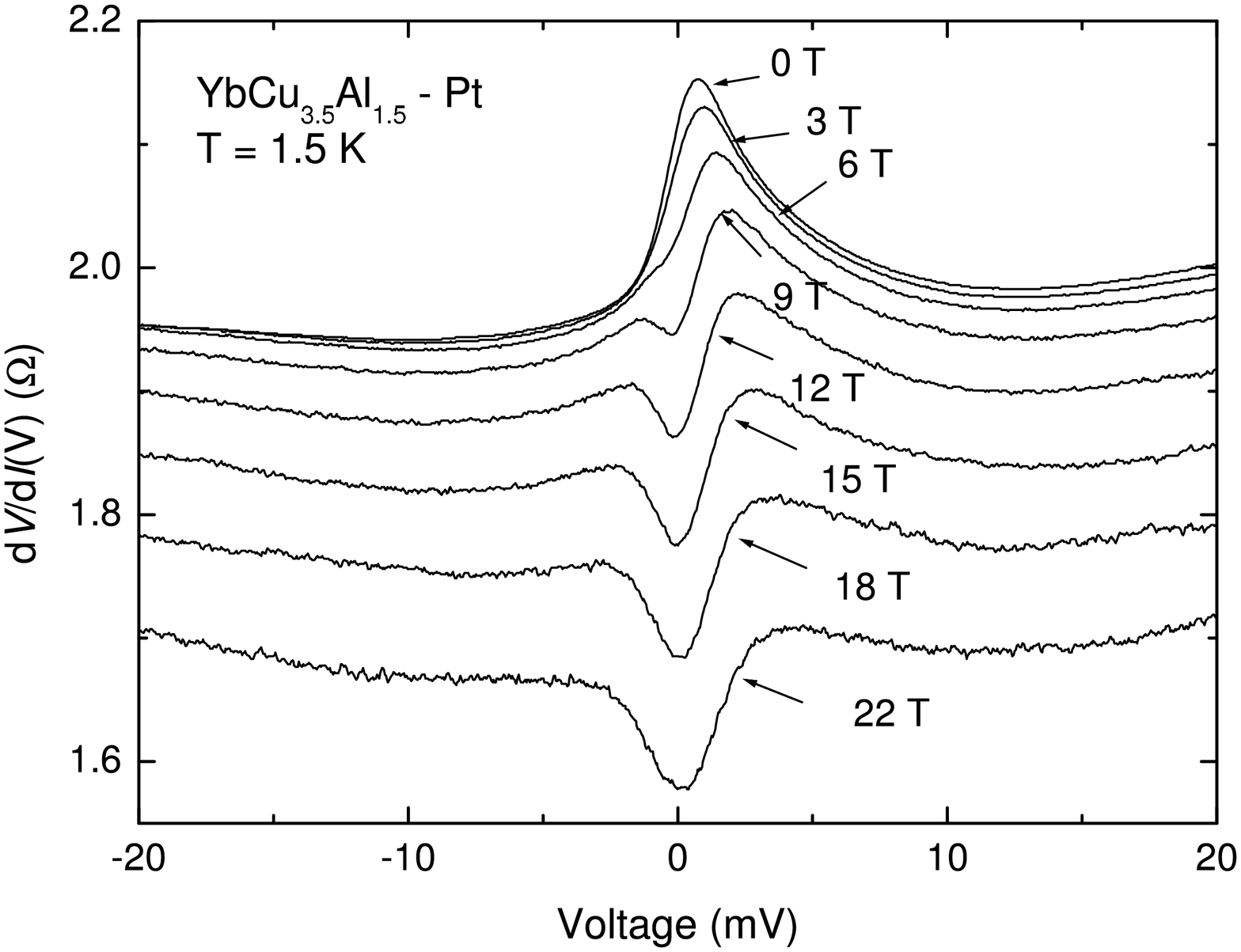}
\caption{\label{fig:YbCu35Al15_B} Characteristic magnetic field behavior of d$V$/d$I$(V) for hetero-contact YbCu$_{3.5}$Al$_{1.5}$ - Pt at 1.5 K.}
\end{figure}

\section{RESULTS AND DISCUSSION}
In order to investigate the influence of a magnetic field, which is
known to destroy the NFL state in the YbCu$_{5-x}$Al$_{x}$ system, we have applied
magnetic fields up to 22.5~T. The d$V$/d$I$(V) curves of a
YbCu$_{3.5}$Al$_{1.5}$ - Pt hetero-contact at 1.5 K are shown in
Fig. \ref{fig:YbCu35Al15_B} in the applied magnetic fields. With
increasing magnetic field the characteristic maximum at 1.3~mV (only
present in one polarity of the applied voltage) shifts to higher
voltages and evolves subsequently into a splitted two-peak
structure. In general, the point-contact resistance is decreasing
with increasing magnetic field. For high magnetic fields (above
12~T) the voltage position of both peaks is symmetric with respect
to zero voltage but their intensities are different. This kind of
slightly asymmetric behavior (B $>$ 12 T) is characteristic for the
point-contact d$V$/d$I$(V) dependencies of heavy-fermion compounds
in a magnetic field. \cite{Naidyuk, Reiffers1992} From measurements of
the electrical resistivity, magnetic susceptibility
and specific heat on bulk samples of YbCu$_{3.5}$Al$_{1.5}$
results, that approximately 12 T is enough to
recover FL behavior and that compounds with concentrations $x <
x_{cr}$ exhibit Kondo-like behavior.\cite{Bauer} Therefore, we
conclude that 12 T is enough to recover the FL behavior of the Kondo
type in our PC spectra. This observation gives an argument for the
NFL origin of our observed asymmetry in magnetic fields below about
12 T.

\begin{figure}
\includegraphics [angle=270, width=80mm]{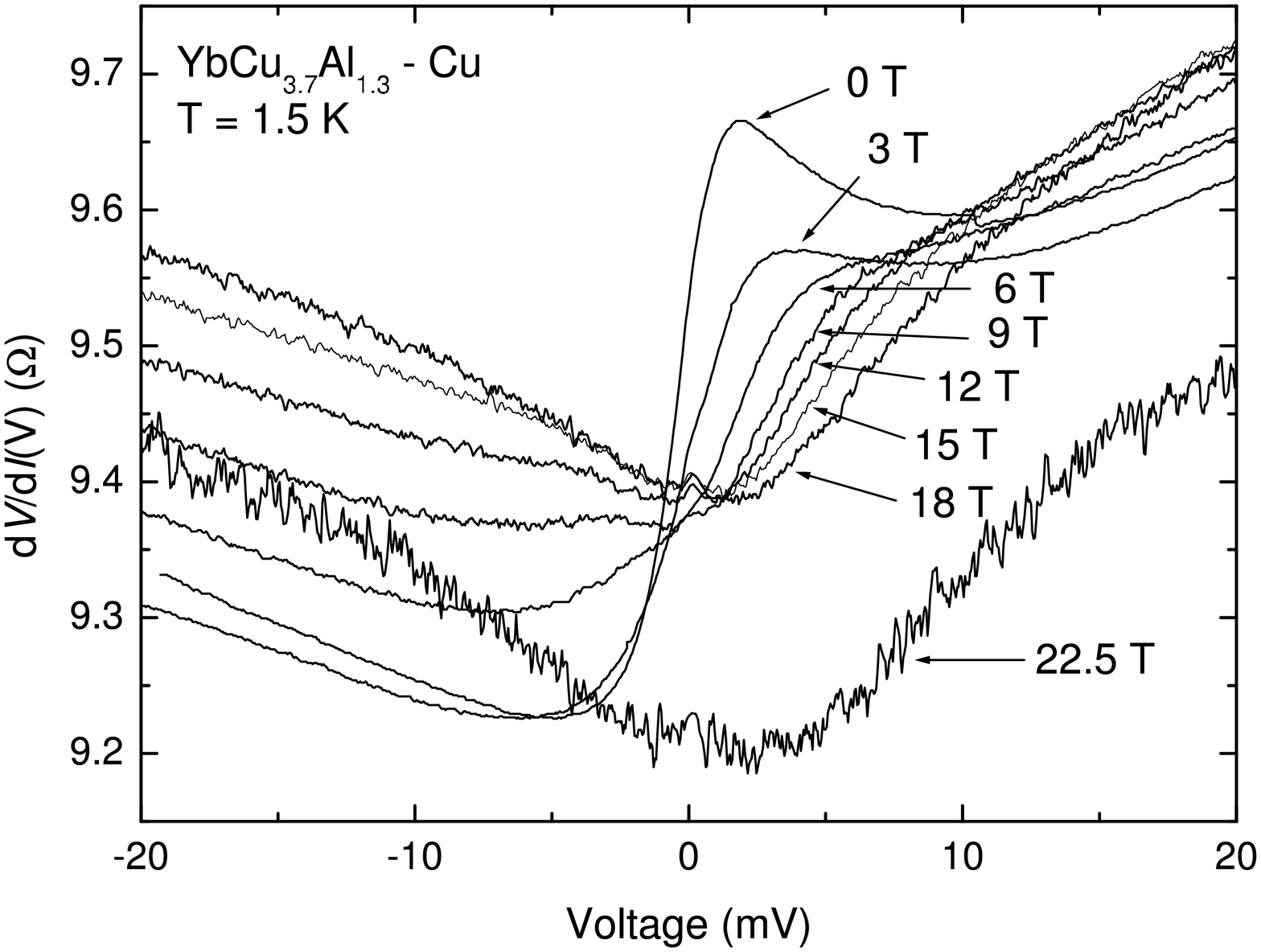}
\caption{\label{fig:YbCu37Al13_B} Characteristic magnetic field behavior of d$V$/d$I$(V) for hetero-contact YbCu$_{3.7}$Al$_{1.3}$ - Cu at 1.5 K.}
\end{figure}

\begin{figure}
\includegraphics [angle=270, width=80mm]{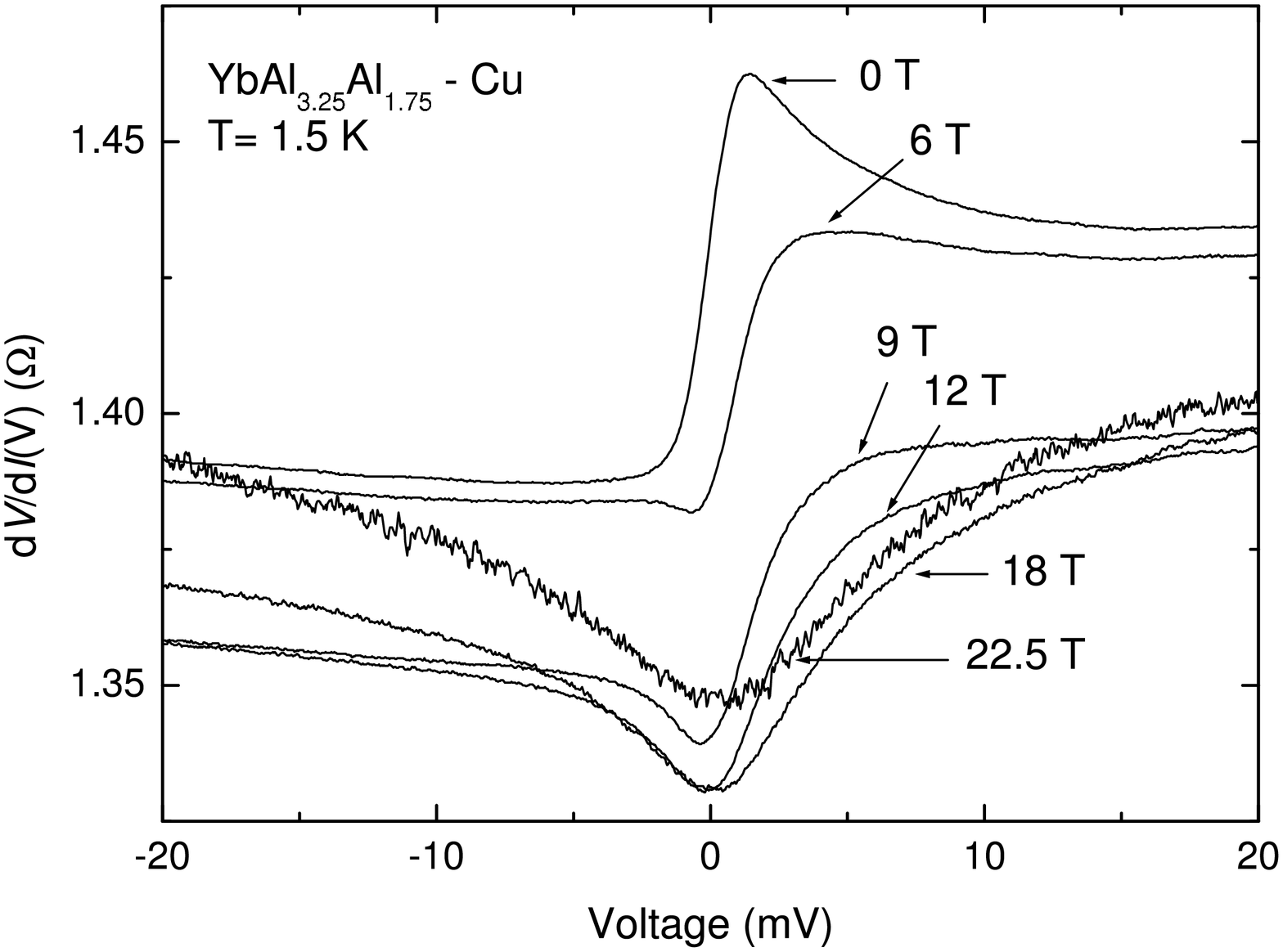}
\caption{\label{fig:YbCu325Al175_B} Characteristic magnetic field behavior of d$V$/d$I$(V) for hetero-contact YbCu$_{3.25}$Al$_{1.75}$ - Cu at 1.5 K.}
\end{figure}

Further confirmation of this conclusion can be seen in the
point-contact characteristics of heterocontacts for $x \neq x_{cr}$
shown in Figs. \ref{fig:YbCu37Al13_B} and \ref{fig:YbCu325Al175_B}.
In comparison with Fig. \ref{fig:YbCu35Al15_B} one can see for $x =
1.3$ and $x = 1.75$ data that the asymmetric maximum is already
suppressed by weaker magnetic fields and the remains of a splitted
two-peak structure are suppressed in the highest magnetic fields.
The main difference for the $x \neq x_{cr}$ data lies in
the high field response which tends to a metallic
like d$V$/d$I$(V) dependence while for $x = x_{cr}$ the maxima in
d$V$/d$I$(V) remain like in PCs of the Kondo type. \cite{Reiffers1992, Duif1989} 

\begin{figure}
\includegraphics [angle=270, width=80mm]{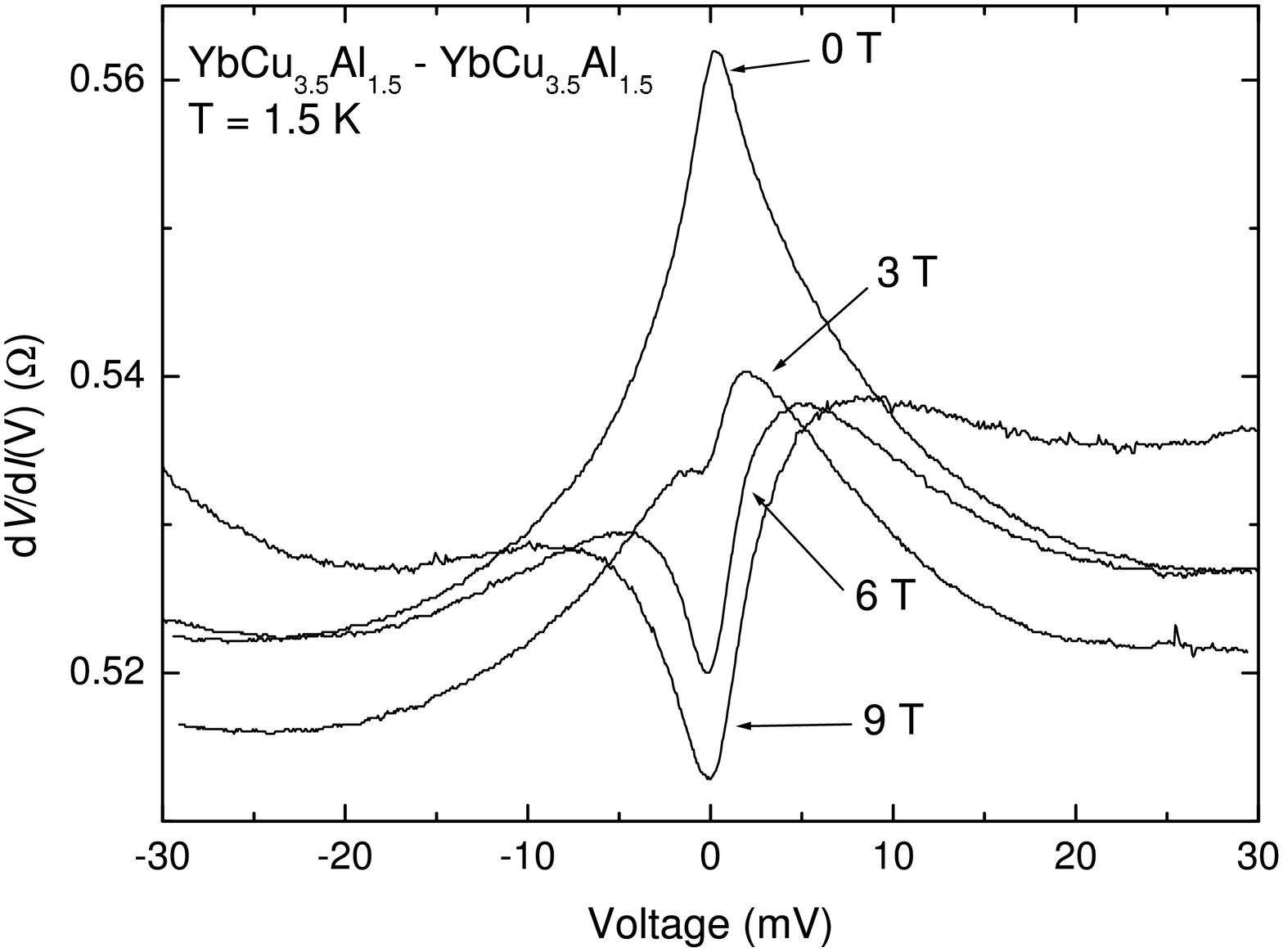}
\caption{\label{fig:hYbCu35Al15_B} Characteristic magnetic field behavior
of d$V$/d$I$(V) for homo-contact YbCu$_{3.5}$Al$_{1.5}$ - YbCu$_{3.5}$Al$_{1.5}$ at 1.5 K.}
\end{figure}

In order to shed light on the origin of the observed asymmetric
maximum in the hetero-contacts, we performed the PC d$V$/d$I$(V)
measurements in the homo-contact arrangement. In Fig.
\ref{fig:hYbCu35Al15_B} the characteristic behavior of homo-contact
is presented (up to 9 T). Because of differences in
the magnetic forces between the contacting parts in the case of the
homo-contact arrangement, the contacts were less stable in an
applied magnetic field compared to the case of
hetero-contacts. The main result is that the maximum in d$V$/d$I$(V)
occurs for the homo-contacts at zero applied voltage. With the
applied magnetic field the splitting of the maximum (symmetrically
positioned around zero voltage) occurs like in case of
hetero-contacts. In the case of homo-contacts for $x \neq x_{cr}$ we
also observed a maximum in d$V$/d$I$ at zero-bias voltage splitting
up in an applied magnetic field.

\begin{figure}
\includegraphics [angle=270, width=80mm]{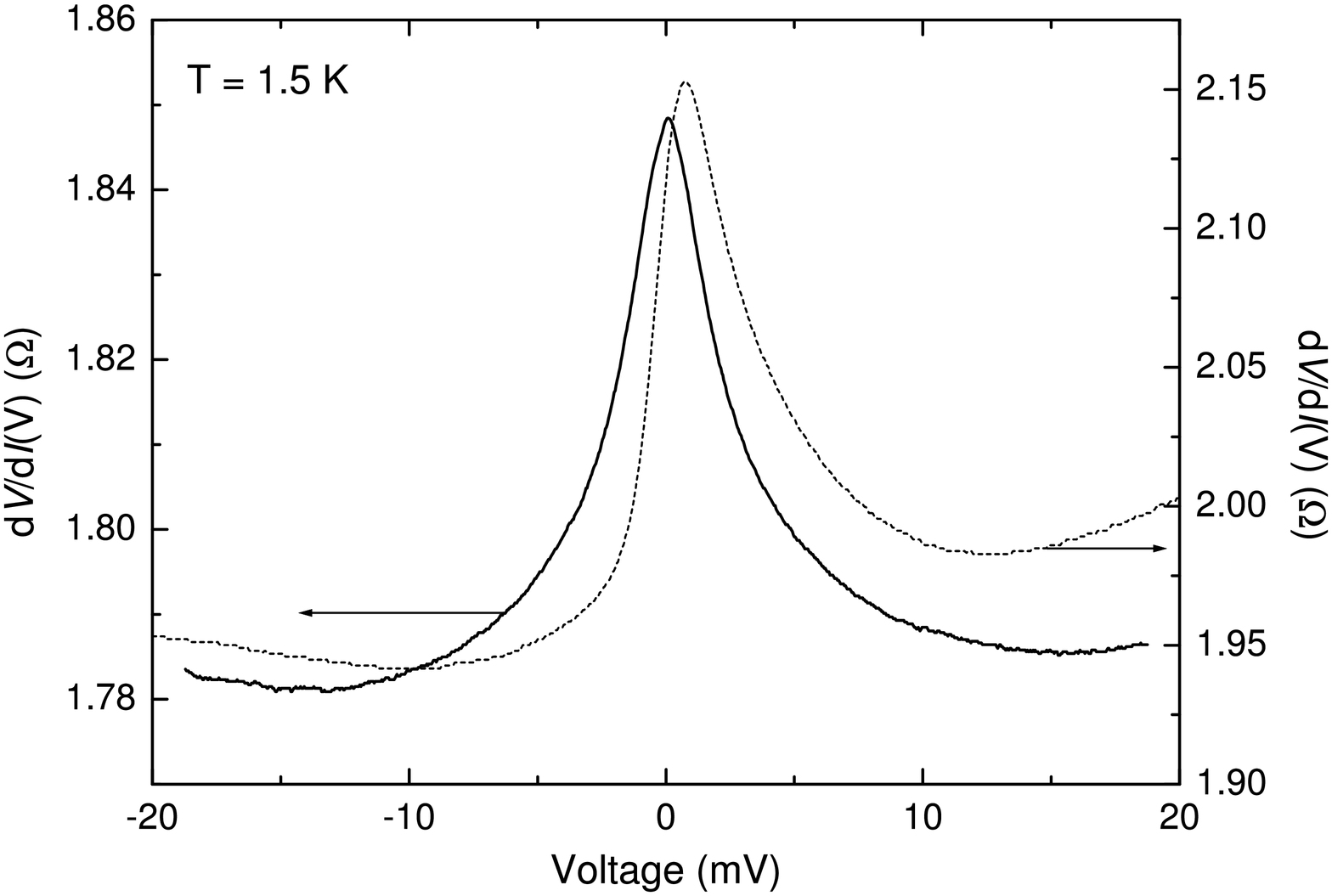}
\caption{\label{fig:homo_vs_hetero} Characteristic d$V$/d$I$(V) for
hetero-contact YbCu$_{3.5}$Al$_{1.5}$ - Pt (dashed line) and
homo-contact YbCu$_{3.5}$Al$_{1.5}$ - YbCu$_{3.5}$Al$_{1.5}$ (solid
line) arrangement at 1.5 K. }
\end{figure}

The comparison of d$V$/d$I$(V) curves  for
homo-contact YbCu$_{3.5}$Al$_{1.5}$ - YbCu$_{3.5}$Al$_{1.5}$ and for
hetero-contact YbCu$_{3.5}$Al$_{1.5}$ - Pt is shown in Fig.
\ref{fig:homo_vs_hetero}. The observed asymmetry in the
current-voltage characteristic does not depend on the used needle
(Cu or Pt). The asymmetry observed in the d$V$/d$I$(V) curves of the
hetero-contact is directly related to the configuration with a
contact between metals with different material properties. From the
similarities between the point-contact spectra of the hetero and
homo contacts, we conclude that the d$V$/d$I$ maximum is connected to
the properties of YbCu$_{5-x}$Al${_x}$ electrode of the contacts.
The asymmetric voltage-position of the maximum is only observed in
the NFL phase of the samples.

In order to explain the observed behavior in the point-contact
spectra we have to consider the regime of current flow across the
constriction, which can be ballistic, diffusive or thermal,
depending on the mean-free-path length, elastic $l_{el}$ and
inelastic $l_{in}$, with respect to the contact diameter $d$.
Covering the whole range of the mean free path values with respect
to the PC diameter, a simple formula was derived by Wexler for the
PC resistance\cite{Wexler}
\begin{equation}
\label{wexler} R_{PC} = \frac{16\rho l_{el}}{3\pi
d}+\beta\frac{\rho(T)}{d}
\end{equation}
where $\rho l_{el}$ = $p_{F}$/$ne^{2}$ with $\rho$ the specific
resistivity, $p_{F}$ the Fermi momentum, $n$ the density of
electrons, and $e$ is the electron charge. The coefficient $\beta$
$\simeq$ $1$ for \textit{l}$_{el}$ $\ll
d$.\cite{PCS} The Wexler formula represents simply an interpolation between the ballistic Sharvin
(\textit{l}$_{el}$ $\gg$ d) resistance (1st term) and the diffusive
Maxwell (\textit{l}$_{el}$ $\ll$ \textit{d}) resistance (2nd term).
In our case the $\rho$(T) of YbCu$_{3.5}$Al$_{1.5}$ is about 100
$\mu\Omega$cm (at T = 4.2 K), and typical values of R$_{PC}$ are
from 0.5 to about 30 $\Omega$. Then from equation (\ref{wexler}), we
can estimate the diameter of the PCs between 50 and 2000 nm taking
the typical value of $\rho l_{el}= 10^{-11}\Omega$cm$^{2}$ for
metals.\cite{PCS} We estimate the typical elastic mean free path,
$l_{el}$, for our compounds in the nm range. As a result, one can
hardly expect ballistic transport of the conduction electrons across
the PC for the $x_{cr}$ compound, but there exists still the
possibility of a long inelastic diffusion length at low applied
voltage as compared to the contact diameter. For an inelastic
diffusion length $(l_{in}l_{el})^{1/2} > d \geq l_{el}$ the
point-contact data can still contain spectroscopic information where
the applied voltage defines the energy scale.

\begin{figure}
\includegraphics [angle=270, width=80mm]{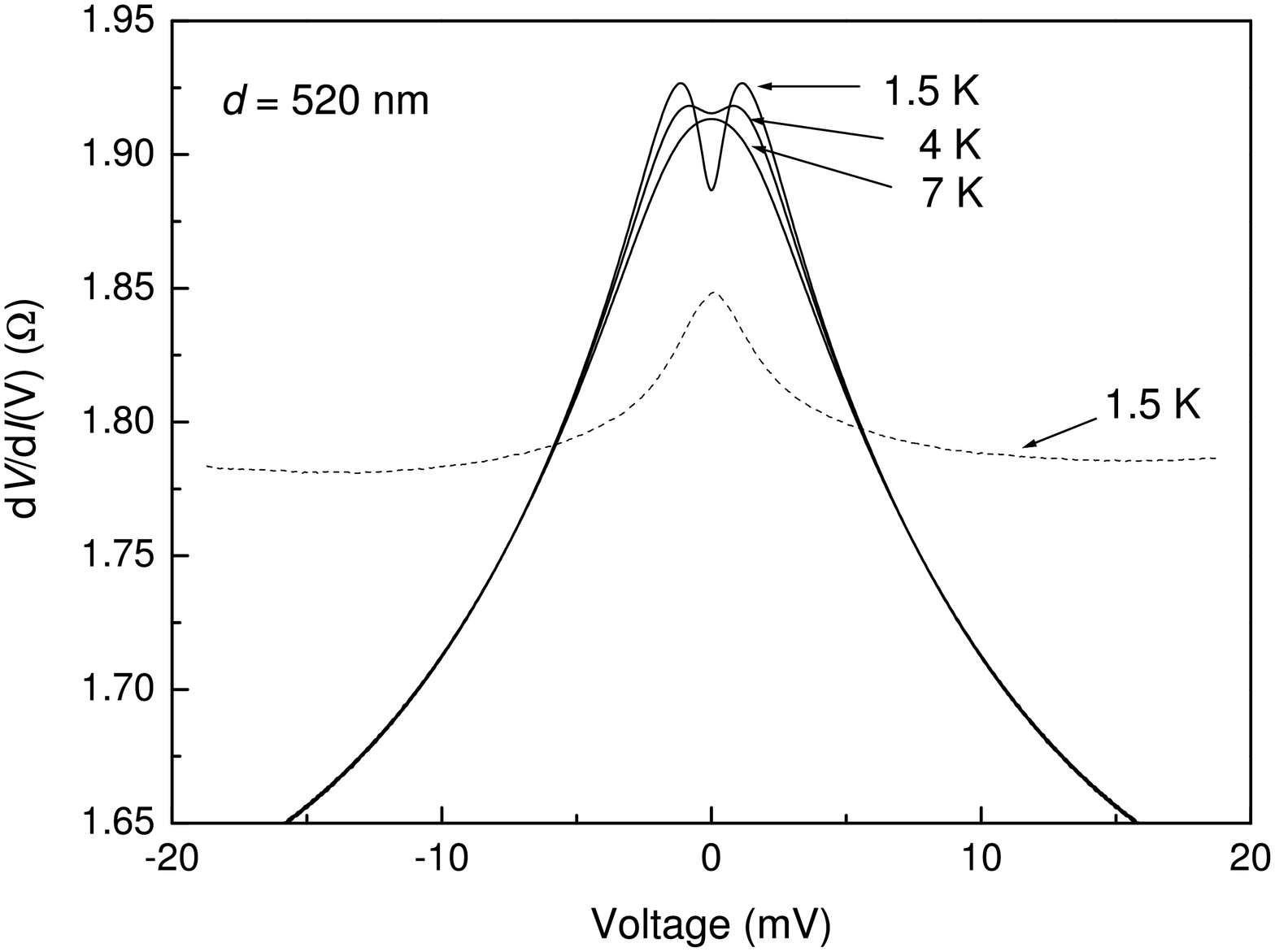}
\caption{\label{fig:homo_model} Comparison of calculated
d$V$/d$I$(V) curves at different temperatures (solid curves) and
experimental data of homo-contact YbCu$_{3.5}$Al$_{1.5}$ -
YbCu$_{3.5}$Al$_{1.5}$ (dashed curve) at 1.5 K.}
\end{figure}

In order to exclude thermal regime where $l_{el},l_{in} < d$ and
energy-resolved spectroscopy is no longer possible, we performed critical analysis of our data. In case of thermal regime d$V$/d$I(V)$
resembles $\rho(T)$ with local contact heating, where a simple
relation exists between the contact resistance as a function of an
applied voltage and the resistivity as a function of temperature
\cite{Kulik}
\begin{equation}
\label{Kulik} I(V)=Vd\int^{0}_{1}\frac{dx}{\rho(T\sqrt{1-x^{2}})},
\end{equation}
with $T = eV/3.36k_B$ for the maximum temperature at
the contact center as a function of the applied voltage.

For the heated contact between two different metals,
the thermoelectric voltage caused by the difference of the Seebeck
coefficients of the contacting metals results in an asymmetry of the
differential resistance curves versus bias voltage. To quantify the
asymmetry of the differential resistance curves, we separate the
d$V$/d$I$(V) curves into the symmetric part d$V$/d$I$(V)$^{s}$ =
[d$V$/d$I$(V$<$0) + d$V$/d$I$(V$>$0)]/2 and the asymmetric part
d$V$/d$I$(V)$^{as}$ = [d$V$/d$I$(V$>$0) - d$V$/d$I$(V$<$0)]/2.
For our experimental configuration, the negative
potential is connected to the YbCu$_{5-x}$Al${_x}$ sample. As was
shown by Itskovich \textit{et al}.\cite{Itskovich},
the asymmetric differential resistance behaves like
\begin{equation}
\label{Itskovich}
\frac{1}{R_{0}}\left(\frac{dV}{dI}(V)\right)^{as}\propto
S_{1}(T_{PC})-S_{2}(T_{PC})
\end{equation}
where $S_{1,2}$ are the Seebeck coefficients of the two metals. The
maximum temperature $T_{PC}$ in the contact is
determined by
\begin{equation}
\label{temp} T_{PC}^{2}=T_{bath}^{2}+\frac{V^{2}}{4L},
\end{equation}
where T$_{bath}$ is the surrounding bath temperature
which reduces to the previously cited expression $T_{PC} =
eV/3.36k_B$ for the free-electron Lorenz number $L$. If the metal 1
has much higher resistivity and thermopower in
comparison to the metal 2, we have d$V$/d$I^s$(V) $\propto$
$\rho_{1}$(T) and d$V$/d$I^{as}$(V) $\propto$ $S_{1}$(T).

\begin{figure}
\includegraphics [angle=0, width=75mm]{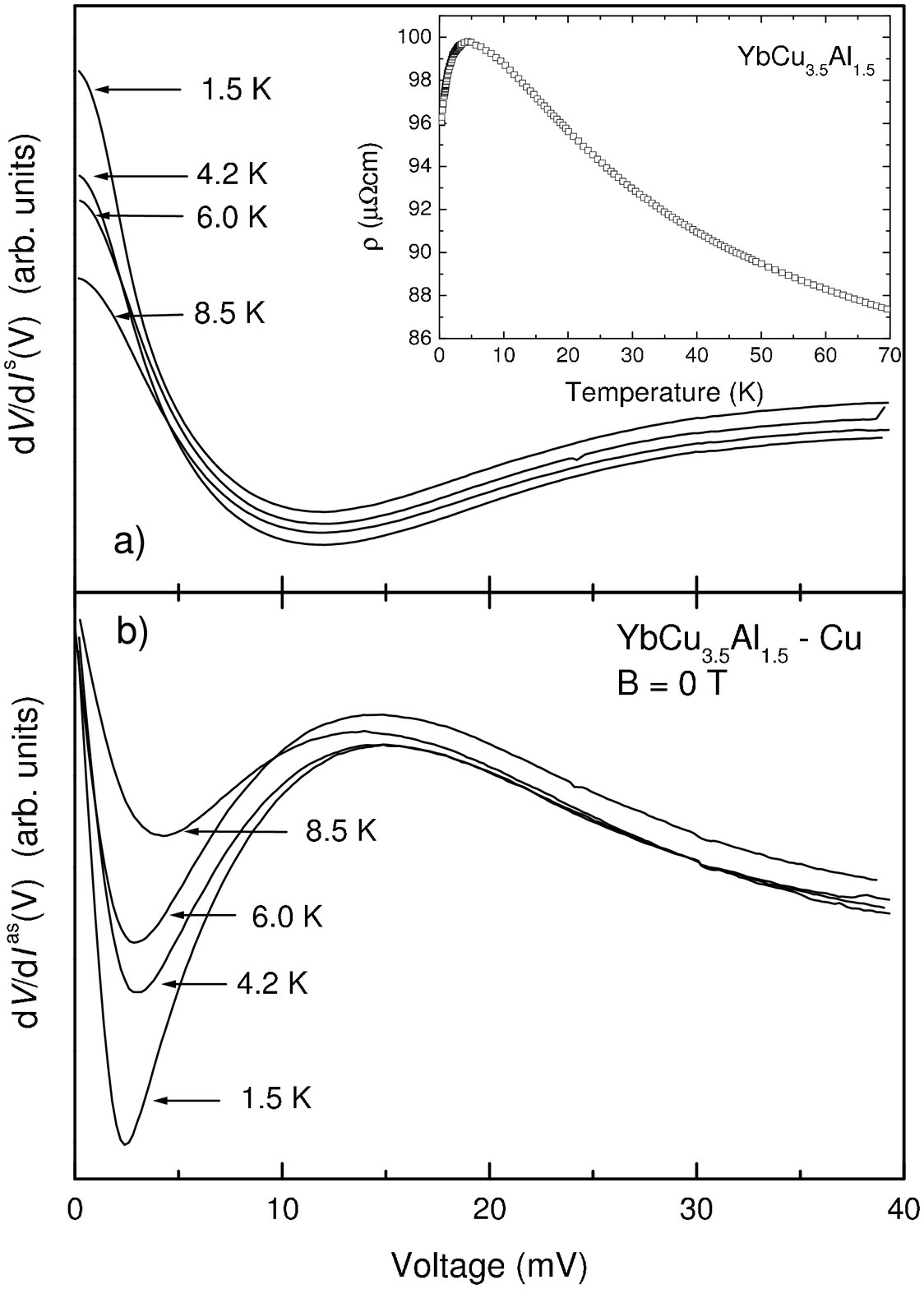}
\caption{\label{fig:sym_asym_T} Characteristic temperature
behavior of symmetric (a) and asymmetric (b) part of d$V$/d$I$(V) for hetero-contact YbCu$_{3.5}$Al$_{1.5}$ - Cu at B = 0 T. The inset shows the bulk resistivity $\rho(T)$ of YbCu$_{3.5}$Al$_{1.5}$.\cite{Bauer}}
\end{figure}

Fig. \ref{fig:homo_model} shows a comparison of the experimental
d$V$/d$I(V)$ data  for a homo-contact of YbCu$_{3.5}$Al$_{1.5}$ at
1.5~K (dashed line) with the calculated characteristics for
different temperatures (solid lines) using the bulk resistivity
data\cite{Bauer} (see also inset in Fig.\ref{fig:sym_asym_T}) in Eq.
(\ref{Kulik}). The diameter $d$ of PC is obtained from the
expression of the Maxwell contact resistance. From the theoretical
analysis, a minimum is expected in d$V$/d$I$ at zero-bias voltage
for temperatures below 4~K corresponding to a decrease in
resistivity in this temperature range. However, in experiment all
the time a zero-bias maximum is observed. In addition, the relative
change in the measured d$V$/d$I$(V) is smaller calculated using Eq.
(\ref{Kulik}). This latter discrepancy between experiment and theory
is often observed in the analysis of point-contact data of the
heavy-fermion compounds and interpreted in terms of an additional
contact resistance resulting from different causes.
\cite{Naidyuk2, Gloos}

Fig. \ref{fig:sym_asym_T} shows the
temperature evolution of symmetric (a) and asymmetric (b) part of
hetero-contact YbCu$_{3.5}$Al$_{1.5}$ - Cu at zero applied magnetic
field. The inset shows the bulk resistivity $\rho(T)$ of this compound.
Also for the case of an hetero-contact, the
behavior of symmetric part d$V$/d$I$(V)$^{s}$ does not show the
corresponding decrease of the bulk $\rho(T)$ (see inset) at the lowest
temperatures as would be expected for the thermal
heating model at the lowest bias voltages.

In order to analyse if experimental broadening could
explain the absence of a minimum in d$V$/d$I$($V$) at zero bias as
expected from the decreasing resistivity at the lowest temperatures,
we will consider the experimental broadening in more detail. The
experimental resolution is given by
$\sqrt{(3.53k_{B}T/e)^{2}+(1.73\sqrt{2}V_{1})^{2}}$, where the first
term results from thermal broadening at a bath temperature $T$ and
the second term from modulation broadening with
amplitude $V_1$ for the recording of first derivative d$V$/d$I$(V)
of the current-voltage characteristic.\cite{PCS} For a typical
modulation amplitude $V_1 = 0.3$~mV, the thermal
broadening dominates at temperatures above about 5~K. In order to
improve thermal resolution and to verify the regime of current
transport trough the contact, we have performed PCS at temperatures
near 100~mK using a dilution refrigerator. Moreover, at these low
temperatures we approach the QCP at $T=0$ as close as possible for
our measurement facility. In the case of $T=100$~mK,
the typically applied modulation voltage of about 0.3~mV determines
the experimental resolution to 0.7~mV.

\begin{figure}
\includegraphics [angle=270, width=80mm]{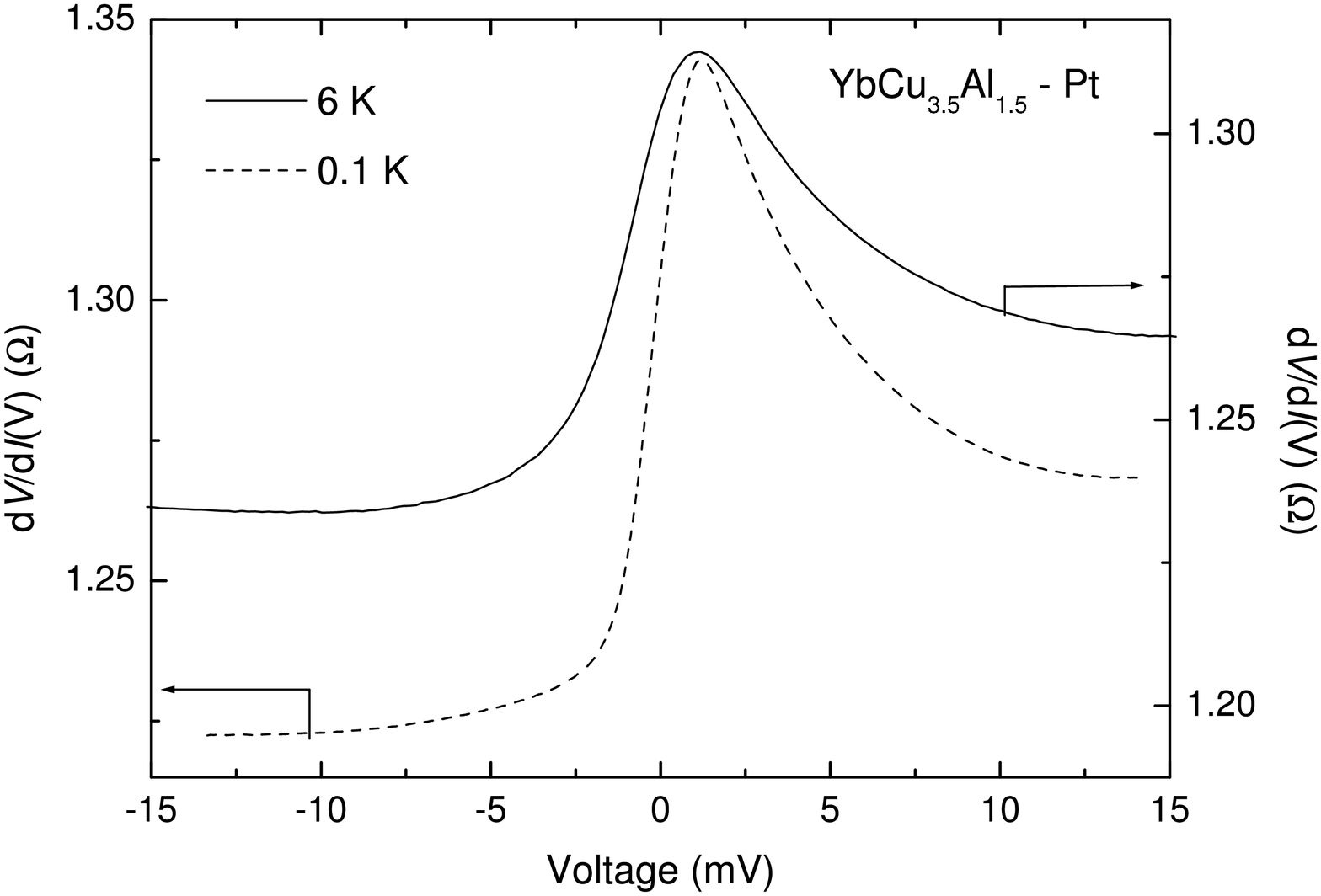}
\caption{\label{fig:comp_100mK_6K} Characteristic d$V$/d$I$(V) of
hetero-contact YbCu$_{3.5}$Al$_{1.5}$ - Pt at 6 K (solid curve) and
at 0.1 K (dashed curve).}
\end{figure}

The comparison of d$V$/d$I$(V) at 6~K and at about 100~mK is
presented in Fig. \ref{fig:comp_100mK_6K}. At lower temperatures the
thermal broadening is reduced, but position of the maximum remains
the same.
Once more, at 100~mK the decrease in $\rho(T)$ at the lowest temperatures
below 4 K (bulk sample) is never observed in d$V$/d$I(V)^s$ at the
lowest bias.

For the case of the thermal regime, the asymmetric parts of the
d$V$/d$I$(V) (see Fig.\ref{fig:sym_asym_T}) would be connected to the
thermopower of YbCu$_{5-x}$Al${_x}$ and we could estimate their
temperature and magnetic field behavior. Taking into account the
work of Mitsuda \textit{et al.} \cite{Mitsuda} and Zlati\'c
\textit{et al.}\cite{Zlatic} for the thermopower of
the related YbCu$_{5-x}$Ag${_x}$ system and other Yb-based systems
and the small magnitude of the Cu thermopower at
low temperatures, one could expect a negative bulk
thermopower for YbCu$_{5-x}$Al${_x}$ with a minimum at low
temperatures. The voltage position of the minimum
(at about 2~mV) would correspond to a temperature
of about 7~K for such a minimum in the thermopower, taking the
free-electron Lorenz number, which value is questionable for NFL compounds.
Moreover, the voltage position of minimum shifts to higher voltages
with increasing temperature, what is in contradiction to the thermal
regime. The shift to lower voltages is expected with its
disappearing when temperature overcome the minimum in thermopower.
Further, the minimum in thermopower of Yb compounds is usually at
about 100 K.\cite{Zlatic} Therefore, this is another argument to exclude the
thermal regime.

\begin{figure}
\includegraphics [angle=0,width=80mm]{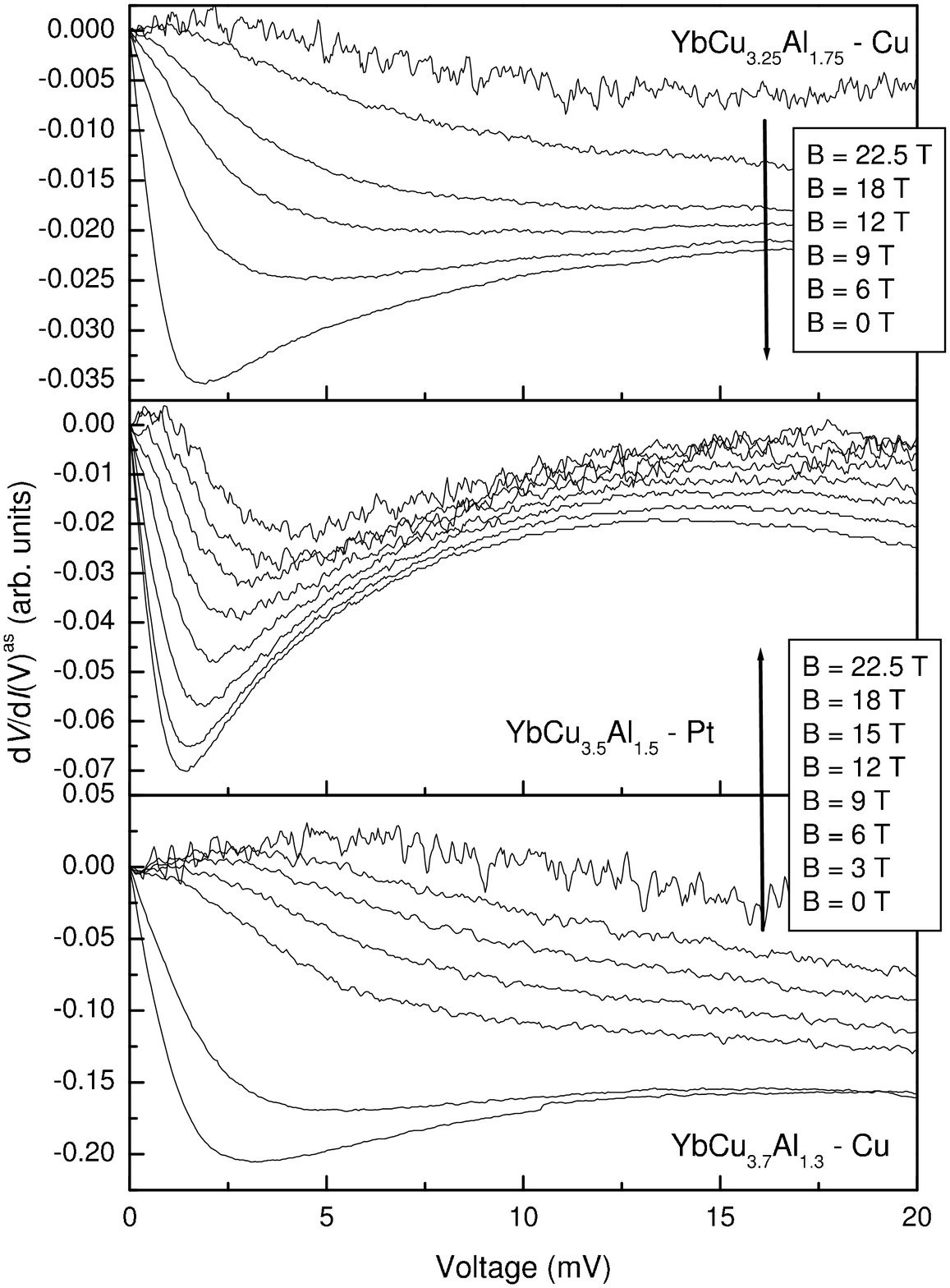}
\caption{\label{fig:asymmetry} Characteristic magnetic field
behavior of d$V$/d$I$(V)$^{as}$ for hetero-contacts with different
x = 1.3, 1.5 and 1.75 at 1.5 K.}
\end{figure}

Fig. \ref{fig:asymmetry} shows the magnetic field influence on the
asymmetric part of d$V$/d$I$(V) for the same hetero-contact
YbCu$_{3.5}$Al$_{1.5}$ - Pt as in Fig. \ref{fig:YbCu35Al15_B}. One
can see that increasing magnetic field suppresses the asymmetry. We
suppose that the application of magnetic field destroys the NFL
behavior and restores the symmetric shape of d$V$/d$I$(V). Taking into account the explanation of the origin of an
asymmetry as connected with the density of states,\cite{Shaginyan2007} we should expect
the linear behavior in the vicinity of zero bias voltage. Our
d$V$/d$I$(V)$^{as}$ dependencies show linear behavior below about 1~mV close
to x$_{cr}$. Then we could suppose that the low voltage behavior of an
asymmetry is governed by this mechanism.\cite{Shaginyan2007} 

This analysis and high magnetic field measurements confirm our previously published data.\cite{Pristas, Reiffers} We have demonstrated that our spectra are not in thermal regime and so they are bringing the information about scattering of conduction electrons. Maximum in differential resistance close to 1~mV reflects increasing magnitude of scattering of conduction electrons. On the other hand the absence of clear maxima in d$^{2}V$/d$I^{2}$(V) in homo-contact as well as in hetero-contact arrangement is in an agreement with concept of the absence of classical quasiparticles in NFL systems.\cite{Stewart, Shaginyan2007} With increasing applied magnetic field we observed features of d$V$/d$I$(V) which are characteristic for Kondo systems. \cite{Reiffers1992, Duif1989} One can see that spectra in NFL regime clearly differs from that of FL regime. New type of d$V$/d$I$(V) asymmetry has been observed, therefore new theoretical explanation is required. We could conclude that the observed new type of asymmetry in the d$V$/d$I$(V) dependencies of hetero-contacts is connected with the
NFL ground state of YbCu$_{5-x}$Al${_x}$ system. The magnetic field
influence on the asymmetry confirms the suppression of the NFL state
and the recovering of the FL state at sufficiently high magnetic fields.

\section{CONCLUSIONS}

The PCS measurements of the NFL compounds
YbCu$_{5-x}$Al${_x}$ in the vicinity of QCP (near $x_{cr}$ = 1.5)
have revealed an asymmetric shape of d$V$/d$I$(V) of the
current-voltage characteristics for the hetero-contact arrangement.
The application of a magnetic field suppresses the asymmetry, alike
the NFL properties. We suppose, that the asymmetry has its origin in
the NFL state of the sample. The analysis of the point-contact differential resistance curves
exclude the thermal regime. We suppose that the asymmetry is governed
by mechanism suggested by Shaginyan and Popov.\cite{Shaginyan2007} Forthcoming thermopower data for the
investigated compound would allow to support experimentally the
observed asymmetry in the NFL part of the phase diagram.

\begin{acknowledgments}
This work has been partly supported by the COST -
ECOM P16, by the Slovak grant agency VEGA 6195, by the Slovak
Academy of Sciences for the Centers of Excellence, by the Science
and Technology Assistance Agency - contract no. APVT-51-031704, and
by the European Commission ($6^{th}$ FP) "Transnational Access -
Specific Support Action" - contract No RITA-CT-2003-505474. US Steel
Ko\v{s}ice sponsored liquid nitrogen needed for the experiments. We
wish to acknowledge the support by the Austrian FWF P18054.
\end{acknowledgments}

\bibliography{references}

\end{document}